\documentclass[preprint,12pt]{elsarticle}

\usepackage{graphics}

\usepackage{amssymb}

\journal{Physica C}

\begin{document}

\begin{frontmatter}



\title{Hysteresis and stepwise  structure in MR curves of
granular superconducting ruthenocuprates
RuSr$_2$(Gd$_{1.5}$Ce$_{0.5})$Cu$_2$O$_{10-\delta}$}

\author{B. I. Belevtsev\corref{cor1}\fnref{label1}}
\ead{belevtsev@ilt.kharkov.ua} \cortext[cor1]{Corresponding
author. Fax: ++380-57-3403370, Phone: ++380-57-3410963}
\author{E. Yu. Beliayev\fnref{label1}}
\author{D. G. Naugle\fnref{label2}}
\ead{naugle@physics.tamu.edu}
\author{K. D. D. Rathnayaka\fnref{label2}}
\address[label1]{B. Verkin Institute for Low Temperature Physics and
Engineering, National Academy of Sciences, Kharkov, 61103,
Ukraine}

\address[label2]{Department of Physics, Texas A\&M University, College Station,
 TX 77843, USA}

\author{}

\address{}

\begin{abstract}
Granular superconductivity effects in a polycrystalline sample of
ruthenocuprate RuSr$_2$(Gd$_{1.5}$Ce$_{0.5})$Cu$_2$O$_{10-\delta}$
are studied. The main attention has been devoted to manifestation
of these effects in current and magnetic-field dependences of
resistive transition to superconducting state. It is found that
current dependences of differential resistance taken at different
temperatures intersect strictly at two definite values of current
demonstrating crossing point effect. This phenomenon has been
explained taking into account inhomogeneous state of intergrain
medium which can be considered as a two-component system. The
particular attention has been given to magnetoresistance (MR)
hysteresis in mixed state of this inhomogeneous system and to
influence of applied current and temperature on this phenomenon.
Two types of hysteresis (clockwise and anticlockwise) have been
found with transition from clockwise to anticlockwise hysteresis
with increasing temperature. Stepwise  structure in MR hysteretic
curves has been observed in low-field range. Possible reasons of
the change in hysteresis behavior with increasing temperature and
appearance of the stepwise structure in MR curves are discussed
taking into consideration inhomogeneous state of the granular
superconductor studied.

\end{abstract}


\end{frontmatter}


\section{Introduction}
Ruthenocuprates of composition
RuSr$_2$R$_{2-x}$Ce$_x$Cu$_2$O$_{10-\delta}$ (Ru1222), with R =
Gd, Eu, have attracted much attention in the last two decades as
certain type of magnetic superconductors (for $0.4\leq x \leq
0.8$) with $T_c$ up to $\approx 50$~K
\cite{felner1,lorenz,awana,klamut}. Superconductivity is
associated with CuO$_2$ planes, while magnetic order is thought to
be connected with the RuO$_2$ planes. The ruthenocuprates show
indications of different magnetically ordered states (from those
that below 80-100 K are thought as being weak ferromagnetic). The
nature and source of the magnetic order in these compounds is
however still not clear. It has been possible so far to prepare
only polycrystalline samples of Ru1222 (by a solid-state reaction
method) \cite{felner1,lorenz,awana}. The samples usually contain
different impurity phases \cite{knee,hata}, so that they are
actually multiphase. The extrinsic disorder and inhomogeneities
depend in a crucial way on the preparation conditions, especially
on sintering and annealing temperature. In some studies, however,
the multiphase state in ruthenocuprates is interpreted as phase
separation effects \cite{lorenz,lorenz2}. In any case, this makes
understandable a variety of magnetic transitions reported in the
literature for this system. A firm belief persists, however, that
magnetic ordering is intrinsic in the primary superconducting
phase in Ru1222 compounds rather than being attributed to some
impurity phases.
\par
Polycrystalline structure of the ruthenocuprates determines the
pronounced granular superconductivity effects in these compounds
\cite{felner2,virg,kras,boris}. The results found are quite
similar to those known for granular high-$T_c$ superconductors,
and can be considered in a similar way. The system of this type
can be described as an ensemble of type-II superconducting grains
(with a size of a few $\mu$m) with weak-coupling intergrain
correlation. Grain boundaries represent weak links (regions of
weakened superconductivity). Intergrain correlation is determined
by Josephson coupling. The following types of weak links are
commonly considered: (i) tunnelling of Cooper pairs across an
electrically insulating boundary interface (SIS junction); (ii)
overlap of superconductive wave functions of the grains in the
normally conducting boundary (SNS, or proximity effect, junction);
(iii) supercurrent transport through tiny superconducting
constrictions (``pinholes'') bridging the grains (a point contact
junction); (iv) SS$^{\prime}$S coupling (S$^{\prime}$ is a
superconductor with lower T$_c$). The coherence length $\xi_{GL}$
in high-$T_c$ superconductors is short and comparable to the grain
boundary thickness. For this reason SIS Josephson junctions are
thought to be of little importance in supercurrent transport while
others of the above-mentioned types of the weak-coupling junctions
are undoubtedly involved in current flowing in granular high-$T_c$
superconductors. Generally, any grain in the superconductor makes
multiple contacts with neighboring grains, so that a bulk
superconductor is a multiple 3D Josephson junction array
\cite{shea}. The intergrain contacts of any grain can be of
different types \cite{hilg}.
\par
Granular superconductors have a widened (in contrast to that in
homogeneous superconductors) temperature range of resistive
transition to the superconducting state, with a ``shouldered''
form of resistive curves $R(T)$, which is also a feature of the
ruthenocuprates \cite{felner2,virg,kras,boris}. The shouldered
form of the resistive transition reflects the inhomogeneous state
of a granular superconductor. With decreasing temperature the
intragrain superconducting transition takes place at temperature
$T_{c0}$, but due to poor intergrain coupling the total resistance
does not go to zero. Only a small drop in resistance can be seen
(if some part of the grains becomes well coupled). In other cases,
a barely noticeable change in $R(T)$ curvature can be noted near
$T=T_{c0}$ \cite{boris}. Intergrain Josephson coupling is
strengthened with further decreasing temperature below $T_{c0}$,
so that in a complicated network of the weak links between grains
some discontinuous zero-resistance paths (or clusters) appear
causing further resistance drop. Due to the inevitable space
distribution of grain boundary thicknesses (and, consequently,
junction resistances) this resistive transition has a percolative
character. As the temperature continues to fall, the
superconducting clusters grow, and (if the space disorder in
junction resistances is not very strong) an infinite percolating
cluster is formed at a certain temperature $T_{c}$ at which the
resistance becomes zero. This temperature can be far less than
$T_{c0}$.
\par
Resistive transition to zero resistance in a weak-link network is
very sensitive to applied current and magnetic field. In granular
high-$T_c$ superconductors, magnetoresistance (MR) curves taken
with increasing and decreasing external field, $H_{ext}$, are
irreversible in the temperature range between $T_{c0}$ and
$T_{c}$. It is found that this MR hysteresis has a peculiar
behavior \cite{qian,dimos,sonin,kopel,ji}, which is quite
different from that in homogeneous superconductors. In the latter
case the resistance in the decreasing $H_{ext}$ curves is higher
than that in the decreasing $H_{ext}$ ones, which is usually
explained by the flux trapping. However, in granular high-$T_c$
superconductors the resistance when the field is decreasing is
lower than that measured with increasing field \cite{kopel,ji}.
\par
The MR hysteresis is closely connected with the magnetic
hysteresis, and both of them reflect variations in the vortex
state of superconducting system in the mixed state with increasing
and decreasing magnetic field. Some models and explanations of
specific MR hysteresis in granular high-$T_c$ superconductors have
appeared in the first years after discovery of superconducting
cuprates \cite{qian,dimos,sonin,kopel,ji}. These have established
some basic approach to consideration of this phenomenon, but due
to complexity of granular systems some questions remain
unanswered. For this reason rather intensive investigations of
this problem continue until recent years
\cite{kilic,balaev,derev}.
\par
MR hysteresis  in the ruthenocuprates in the range of resistive
superconducting transition has been noticed and discussed in
\cite{felner2,virg}. In this article further study of this
phenomenon in the ruthenocuprates is presented. The Gd doped
sample RuSr$_2$Gd$_{1.5}$Ce$_{0.5}$Cu$_2$O$_{10-\delta}$
(Ru1222-Gd) has been studied. The study is to some degree a
continuation of our previous studies \cite{kras,boris} of granular
superconductivity effects in the Ru1222-Gd. We will describe here
some new results not mentioned in previous articles
\cite{kras,boris}. In particular, a stepwise structure in MR
curves taken with increasing and decreasing field has been found
and discussed.

\section{Samples and experimental technique}

The Ru1222-Gd samples studied have been prepared by a solid-state
reaction method in I. Felner's lab \cite{felner1}. Some of them
were set aside (as-prepared samples), while others were annealed
(12~h at 845~$^{\circ}$C) in pure oxygen at different pressures.
It was shown previously \cite{boris,boris2} that oxygen annealing
of these samples has a rather strong effect on intergrain
connection but does not influence intragrain superconducting
properties. This study focuses on properties of as-prepared
Ru1222-Gd samples where the granularity effects were most evident.
Resistance as a function of temperature and magnetic field (up to
16 kOe) was measured using a standard four-point probe technique
in a home-made cryostat. The magnetization measurements were made
with a Quantum Design SQUID magnetometer. The samples were
polycrystalline with a grain size of a few $\mu$m. They were
characterized by resistivity, magnetization and specific heat
measurements, which were in part reported in \cite{boris,boris2}.

\section{Results and discussion}

\subsection{General characterization of the sample}
Some general characteristics of transport properties the sample
studied will be considered before going to the main results about
MR behavior in the range of the resistive superconducting
transition. Temperature dependences of the resistance $R(T)$ of
the as-prepared sample
RuSr$_2$(Gd$_{1.5}$Ce$_{0.5}$)Cu$_2$O$_{10-\delta}$ have been
recorded for different measuring currents in the range 0.01--50
mA. Fig. 1 shows $R(T)$ dependences only for currents at which MR
hysteresis was studied. The temperature $T_{c0}=34$~K (shown by
arrow) marks a kink in the $R(T)$ dependences and is attributed to
the intragrain superconducting transition. This value of $T_{c0}$
is determined on the basis of transport, magnetic and heat
capacity measurements \cite{boris,boris2}. Weak intergrain
connection causes nonmetallic behavior of $R(T)$ above $T_{c0}$
($dR/dT <0$) with an approximately logarithmic law ($\Delta R
\propto \ln T$). Below $T_{c0}$ resistance continues to increase
with decreasing temperature but not so rapidly and only up to
temperature $T_{cJ}=23.2$~K at which the maxima in $R(T)$ curves
for all measuring currents takes place (Fig. 1). This presents
also the branching point of the family of $R(T)$ curves taken for
different currents. The temperature $T_{cJ}$ indicates a
temperature point below which percolating chains of intergrain
Josephson coupling are created, and the resistance starts to
decrease with decreasing temperature. Due to strong disorder in
the intergrain coupling, the resistance in the sample studied does
not go to zero with decrease in temperature down to 5 K (Fig. 1).
\par
General interpretation of this $R(T)$ behavior and influence of
measuring current on it can be found in \cite{boris}. As the
measuring current is increased, some of the weakest links of the
percolative chains with the least critical current go to the
resistive state, increasing the total resistance of the system. A
similar action (and for the same reasons) is produced by an
applied magnetic field in the range below $T_{cJ}$ (Figs. 2 and
3). It can be seen that resistance is very sensitive to magnetic
field especially within the low field range. MR curves show no
saturation with increasing field except for temperatures very
close to $T_{c0}=34$ K (Fig. 3) when total superconductivity
suppression is possible at rather low field.
\par
Characterization of ruthenocuprates is incomplete without
magnetization data. Most important results concerning this and
similar Ru1222 samples are already presented in previous papers
\cite{boris,boris2}. Here we would like to show only typical
magnetization hysteresis curve in the temperature range of
superconducting state (see Fig. 4 for the Gd ruthenocuprates),
which agree with known data \cite{hata}. Below $T_c$ the Ru1222
compounds show a ferromagnetic-like behavior together with
diamagnetic (Meissner) response at low applied fields
\cite{hata,boris,boris2}. At higher fields no saturation takes
place due to paramagnetic contribution from rare-earth ions or
impurity phases (Fig.~4). Changes in $M(H)$ curves with increasing
temperature are shown in Fig.~5. It is seen that nonlinear
low-field $M(H)$ behavior at low temperatures changes to linear at
higher temperatures.
\par
Current-voltage characteristics of the sample studied are strongly
non-linear below $T_{cJ}$ but become approximately linear when
approaching $T_{cJ}$ (and above it). This is seen in Fig.~6 where
current dependences of differential resistance $R_{dif}=dU/dJ$ are
shown at different temperatures. The curves $R_{dif}(J)$ have two
characteristics crossing points (the first on the ascending parts
of the curves and the second on the descending part). The current
dependences of dc resistance $R=U/J$ (nor shown) have the same
appearance as in Fig. 6, but with higher values of the crossing
point currents $J_{x1}$ and $J_{x2}$ (1.95 mA and 35,2 mA,
respectively). At temperatures rather far below $T_{cJ}$ both,
$R_{dif}(J)$ and $R(J)$, curves show weak current dependence for
small currents, but with increasing current the curves go up and
then down forming the maximum (Fig. 6).
\par
The crossing point phenomenon is well known in strongly correlated
electron systems \cite{voll}. It implies that a family of curves
$P(x,y)$ (where $P$ is some physical characteristic, $x$ and $y$
are some thermodynamic or dynamic quantities) can intersect
strictly at one point when $P(x,y)$ is plotted as a function of
one of the variable for different values of the other one. This
was found, for example, for  Gd ruthenocuprates in the
specific-heat temperature and magnetic field curves \cite{boris2}.
Both crossing points in Fig. 6 have the typical appearance for
this phenomenon \cite{voll}. At the same time the two crossing
points observed have some clear individual features. It is seen in
Fig. 6 that only curves taken at $T\leq 15$~K show a crossing
point effect. It is apparent that the first crossing point (at
$J=J_{x1}$) is simultaneously the inflection point for each curves
intersected. This means that at this point
\begin{equation}
\frac{\partial^{2}R_{dif}(J,T)}{\partial J^2}\bigg |_{J_{\ast}(T)}
=0.
\end{equation}
We shall try to explain the non-monotonic behavior of $R_{dif}(J)$
curves and origins for appearance of the crossing points below.
Here we would like to stress mention that the crossing point
effect is a characteristic feature of inhomogeneous (for example,
two-level, two-component, two-phase and the like) systems. It is
asserted \cite{voll}, among other suggestions, that the crossing
point should become apparent in a system which is a superposition
of two (or more) components, like that in the known
Gorter-–Casimir two-fluid model of superconductivity. Granular
superconductors can be considered as some type of two-component or
two-level system as will be discussed below.

\subsection{Hysteresis and stepwise structure in MR curves}
We have studied the magnetic field dependences of the resistance
and MR hysteresis for different values of the transport current
(indicated in Fig.1). The $R(H)$ curves were recorded under
continuous variation of the field from -0.005 T to zero and then
to the field $H_{max}$ (about 1.5 T) with subsequent decreasing to
zero on going to the starting value of -0.005 T. During such
cycles the same sweep rate of 20~mT/s with the field perpendicular
to the transport current is kept. No significant difference in
$R(H)$ curves (including the stepwise structure described below)
has been found by us for measurements under other orientations of
$H$ (for example, parallel or antiparallel to the current). We
have found that the MR hysteresis pattern changes with increasing
temperature (in the range below $T_{cJ}$) as shown in Fig. 7. At
low enough temperature, the MR curves measured with decreasing
field go below those taken with increasing field (clockwise
hysteresis). When temperature increases sufficiently close to
$T_{cJ}$ the opposite behavior takes place: the decreasing $H$
curve goes above that taken with increasing field (anticlockwise
hysteresis). This change in the MR hysteresis behavior with
temperature is found for the transport currents in the range
0.1--3~mA. It is seen that under the clockwise MR hysteresis the
MR curves cross in the low field range (Fig. 7). Similar crossing
has been found in MR hysteresis of high-$T_c$ superconductors with
weak intergrain coupling \cite{balaev}. For both types of
hysteresis the MR is finite (''remanent`` MR) when the external
magnetic field decreases to zero going to the end of the cycle. It
follows from Fig. 7 and those presented below (Figs. 8, 9, and 10)
that the minimum (zero) value of MR with the backward $H$ sweeping
is achieved not at $H=0$ but at some negative value of $H$.
\par
Previous studies of MR hysteresis in superconducting
ruthenocuprates are scanty. Actually it is possible to compare
properly the results obtained only with those in Ref.
\cite{felner2}, where Ru1222-Gd with somewhat different
composition [RuSr$_2$(Gd$_{1.3}$Ce$_{0.7}$)Cu$_2$O$_{10-\delta}$]
and higher $T_{c0}\approx 43$~K has been studied. In that study
the clockwise hysteresis at temperature rather far below $T_{c0}$
and ''remanent`` MR  are found (same as in this study). In
distinction to Ref. \cite{felner2} we have found the transition
from the clockwise to anticlockwise MR hysteresis with increasing
temperature in the range below $T_{cJ}$. Moreover, we have found a
stepwise structure in the $R(H)$ curves at low-field range. This
structure can be seen, for example, if any curve in Fig. 7
(obtained for $J=3$~mA) is enlarged sufficiently, like that shown
in Fig. 8. Similar stepwise structure is found to appear for other
transport currents (Figs. 9 and 10). This structure appears for
both, the upward and downward magnetic-field variations. It is
significant that in both cases the stepwise structure appears only
below some characteristic field $H_x$ (Figs. 9 and 10).
Temperature dependences of the field $H_x$ together with those of
MR at $H=1.6$~T are shown in Fig. 11 at two currents, 0.2 and 0.5
mA. It is evident  that: (i) clear correlation exists between the
temperature dependences of $H_x$ and MR; (ii) $H_x$ goes to zero
with temperature going to $T_{cJ}$. The latter means  that the
stepwise structure appears in MR curves only below $T_{cJ}$.
\par
In the range $T_{cJ}\leq T < T_{c0}$, MR hysteresis is
anticlockwise and without the stepwise structure (Fig. 12). Here
two main features can be observed. First, negative MR in the
low-field range, and, second, saturation of resistance at high
enough field when temperature is fairly close to $T_{c0}=34$~K.
Above $T_{cJ}$ intergrain Josephson coupling is nearly suppressed
so that intergrain conductivity is determined mainly by
single-particle tunneling. In this case negative MR appears
\cite{belufn}. It is associated with the reduction of the
intragrain superconducting gap $\Delta (T )$ in an applied
magnetic field, that causes an increase in the unpaired
quasi-particle density and corresponding decrease in resistance
\cite{belufn}. Although the intragrain upper critical field is
found to be very large in the Ru1222-–Gd compound \cite{kras},
close enough to $T_{c0}$, the field applied in this study (up to
1.6 T) is quite enough to suppress intragrain superconductivity.
It is seen in Fig. 12 that at $T = 32.31$~K the $R(H)/R(0)$
dependence shows negative MR for low field; then with further
increase of the field MR becomes positive and saturates at $H>
0.6$~T. Above this field the resistance is constant implying total
suppression of the intragrain superconductivity by the magnetic
field. MR remains positive when the field is decreased after the
maximal applied field was reached (Fig. 12).

\subsection{Discussion}
The ruthenocuprates are magnetic superconductors but the internal
magnetic field from the spontaneous magnetization in the
Ru1222-–Gd is rather small, below 0.004 T \cite{leviev}. So that,
in the first approximation, this will be not taken into account in
our discussion.
\par
For homogeneous superconductors the anticlockwise hysteresis of MR
(when $R(H)$ under decreasing $H$ goes above that taken with
increasing field) is usually explained by the flux trapping.
Granular superconductors are inhomogeneous and just this
determines appearance of the clockwise behavior of MR hysteresis
in these materials. At present time in known literature  this
behavior is commonly explained on the basis of the two-level
critical-state model \cite{ji} developed for high-$T_c$ granular
superconductors. This model describes actually the two-component
system which consists of (i)  superconducting grains with critical
current $J_{cg}$ and the lower and upper critical fields $H_{c1g}$
and $H_{c2g}$; and (ii) so called, intergrain Josephson medium
with critical current and fields $J_{cJ}$, $H_{c1J}$ and
$H_{c2J}$. It is reasonably taken that $H_{c2g}\gg H_{c2J}$ and
$J_{cg}\gg J_{cJ}$.  The field $H_{c2g}$ in Ru1222-–Gd compounds
is enormously large. For example, it was found that $T_{c0}$
decreases by only about 2 K with increasing field up to 8 T
\cite{boris}. Lower critical field of the intergranular medium,
$H_{c1J}$, is expected to be very small (below $10^{-4}$ T
according to the estimate in Ref. \cite{sonin}). The intragrain
lower critical field, $H_{c1g}$, in ruthenocuprates is poorly
studied, but judging from available \cite{awana,virg} and our own
data its values can be in the range 0.01--0.1 T. It follows that
even at the lowest external field the intergranular medium of the
sample studied is in the mixed state (even with neglect of the
inner magnetic field due to the spontaneous magnetization).
\par
Resistivity in the mixed state of such an inhomogeneous system is
determined primarily by intergrain coupling. Intergrain Josephson
medium is sensitive even to small variations in current and
magnetic field; whereas, these variations (at $J\leq J_{cg}$ and
$H\leq H_{c1g}$) have no influence on intragrain superconducting
properties. With increasing field, the flux penetrates primarily
grain boundary regions with weakened superconductivity causing
depression of intergrain coupling and a corresponding strong
resistance rise in low field (Fig. 3). With further increasing
field (above $H_{c1g}$) the flux begins to penetrate into grains.
It is evident that increase in current exerts a similar influence
on the intergranular medium and, thus,  on measured resistance
(see Fig. 1 at $J\leq J_{cg}$). The relevant parameter which
determines behavior of resistance in magnetic field is the
magnetic flux density (induction) $B_J$ of the intergrain
Josephson medium. A clear correlation between $B_J$ and resistance
in the mixed state of granular high-$T_c$ superconductors was
demonstrated in Ref. \cite{kopel}.
\par
It is generally accepted  that based on the two-level
critical-state model \cite{ji} the intergrain induction $B_J$ is
determined by three contributions: (1) the applied magnetic field,
$H_{ap}$; (2) the intragrain magnetization, $M_g$, and (3) the
intergrain magnetization, $M_J$. Due to complexity of such
systems, in known literature only some average expressions for
$B_J$ \cite{kopel,balaev,mahel,mune} have been presented, which
take into account quite generally sample shape, demagnetization
fields of the sample and the grains, shape and volume fraction of
grains and another factors. This can be written in a simplified
form, for example, as
\begin{equation}
B_J = H_{ap}+M_J(H_{ap},J) -M_g(H_{ap})\,C(H_{ap}),
\end{equation}
where $C(H_{ap})$ is a shape dependent numerical factor. It is
seen from Eq. (2) that $M_J$ and $M_g$ have opposite contributions
to the intergranular magnetic flux density, that has been clearly
shown in \cite{mahel}. Due to weakened superconductivity of the
intergranular medium the relation $M_J\ll M_g$ is expected to
hold, so that Eq. (2) can be in some approximation rewritten as
\begin{equation}
B_J \approx H_{ap} -M_g(H_{ap})\,C(H_{ap}).
\end{equation}
It is seen that hysteresis in MR in granular superconductors is
determined primarily by hysteretic behavior of the intragrain
magnetization, which is characterized by the intragranular flux
trapping. In superconductors, the dependence of $M(H)$ with
decreasing field goes above that with increasing field due to flux
trapping. This leads to the clockwise hysteresis in MR in granular
superconductors. The same phenomenon (flux trapping) determines
the anticlockwise MR hysteresis in homogeneous superconductors
since in this case the common relation $B(H_{ap})=H_{ap} +
M(H_{ap})$ holds. In this study the transition from clockwise to
anticlockwise MR hysteresis with increasing temperature has been
found (Fig. 7).  The anticlockwise hysteresis is fully developed
when temperature is close enough to the characteristic temperature
$T_{cJ}=23.2$~K above which a Josephson-like intergrain coupling
is depressed although grains remain superconducting up to
$T_{c0}=34$ K. In the range $T_{cJ} \leq T \leq T_{c0}$ the
anticlockwise hysteresis takes place (Fig. 12). It is evident that
when $T$ approaches and exceeds $T_{cJ}$ the granular system
studied does not correspond to the two-level critical-state model
\cite{ji} since the intergranular Josephson medium disappears and
transition to one-component system takes place so that MR
hysteresis becomes anticlockwise.
\par
Now we will focus again on the stepwise structure of MR curves
(Figs. 8, 9 and~10).  Some stepwise structure in MR curves has
been reported previously for a high-$T_c$ YBCO polycrystalline
sample \cite{kilic}, but only very general suggestions about this
phenomenon were given. Based on results of this study the
following general features found for the phenomenon can be
distinguished. {\it First}, the stepwise structure appears only
below $T_{cJ}=23.2$~K that is in the region of the assumed
two-level (or two-component) state of the granular superconductor
studied. It can be suggested therefore that the jumps of
resistance in magnetic field are associated with peculiarities of
the flux moving in the intergranular Josephson medium. {\it
Second}, the steps appear in $R(H)$ curves beginning from lowest
applied fields and disappear at some characteristic field $H_x$,
which depends on temperature and applied current and clearly
correlates with MR (Fig. 11). {\it Third}, the relative amplitude
of the resistance jumps or steps ($\Delta R/R_{0}$, where $R_0$ is
resistance in zero field) generally decreases with increasing
field. This is especially evident for higher applied current (see
Fig. 8 for $J=3$~mA). For low currents this is not so obvious
(Figs. 9 and 10). The average amplitude of $\Delta R/R_{0}$
decreases with increasing current and temperature as can be seen
in Fig. 13. This Figure (and Fig.~11) shows that the stepwise
structure exists only below $T_{cJ}$.
\par
The stepwise structure of the MR curves suggests that penetration
of magnetic field into the intergranular Josephson medium proceeds
not smoothly, but discontinuously, step-by-step. This is somewhat
similar (but not the same) to the phenomenon of flux jumps (or
magnetic instability) known for a long time in homogeneous type-II
superconductors \cite{wipf,mints,tinkham}. In certain
circumstances flux penetrates superconductors via discrete jumps
or avalanches where a few or many vortices hop at once from one
position to another \cite{tinkham}. Typically, flux jumps have
revealed themselves as abrupt jumps in the magnetostriction and
magnetization hysteresis loops in the Meissner state. This
phenomenon in homogeneous superconductors shows itself only in
samples of sufficient size (critical dimension criterion)
\cite{tinkham}. The magnetic instability was observed in
high-$T_c$ superconductors as well \cite{htsc}.
\par
Due to the critical dimension criterion, the flux jumps in
high-$T_c$ superconductors have been observed only in rather large
single crystals or well textured polycrystalline samples with high
critical current \cite{htsc}. In ceramics the critical dimension
is determined by grain size which is usually too small for
occurrence of the flux jumps. As far as we know no such effect
(like jumps in magnetization loops) was seen in polycrystalline
ruthenocuprates, so that jumps in intragrain magnetization (which
makes the dominant contribution to global measured magnetization)
can be excluded. But below $T_{cJ}$ the measured resistivity of
the granular superconductor studied  is determined primarily by
intergrain weak links and properties of the intergranular
Josephson medium. This medium is inhomogeneous, so that its
critical parameters such as $J_{cJ}$ and $H_{c2J}$ (and, hence,
intergrain Josephson coupling) have some spatial distribution. As
a result of this, the increase in resistance with increasing
magnetic field cannot proceed smoothly but step-by-step due to
local flux jumps in the Josephson medium causing resistance jumps.
\par
Resistance jumps in MR curves (Figs. 8, 9 and 10) can also be
considered from the point of view which takes into account the
percolative character of intergrain conductivity below $T_{cJ}$.
In a percolative granular system the conductivity is determined by
the presence of ``optimal'' chains of grains with maximum
probability of electron transport for adjacent pairs of grains
forming the chain. As the applied field is increased, some of the
weakest links of the chains go to a resistive state causing an
increase in total resistance of the system. This process can go in
steps as a discontinuous transition from less resistive to a more
resistive set of ``optimal'' chains of grains with increasing
field. It is evident that increasing field leads to a decrease in
the total number of weak links at which superconducting electrons
can transfer from one grain to another. The same influence is
exerted by increasing temperature. In both cases, this means
decreasing in area of the Josephson medium. For high enough field,
the number of weak links (or volume of the Josephson area) becomes
too small to produce noticeable steps in total resistance with
increasing field, and the MR curves become smooth. For this reason
the stepwise structure disappears above a characteristic field
$H_x$ (Figs. 8, 9 and 10) and this field decreases with
approaching to $T_{cJ}$ (Fig. 11). In a resistive state of the
percolative superconducting granular system even a rather low
magnetic field can influence the Josephson medium and cause
resistance jumps as observed in this study. This is in contrast to
flux jumps in homogeneous superconductors where flux jumps are
possible only above the threshold field \cite{wipf}.
\par
Now we can return to discussion of the crossing point effect in
the family of $R_{dif}(J,T)$ curves (Fig. 6), taking into account
that the intergranular  medium which determines the total
resistance below $T_{cJ}$ is inhomogeneous. It consists of (1)
Josephson medium which presents weak links for intergranular
propagation of superconducting electrons, and (2)
nonsuperconducting (normal) medium  through which only transfer of
unpaired quasi-particle excitations is possible. A somewhat
similar concept has been applied previously for low-$T_c$ granular
superconductors \cite{belufn} where resistive transitions were
explained by concurrent tunnelling of unpaired quasi-particle
excitations and intergrain Josephson tunnelling of Cooper pairs.
The intergranular medium is a typical two-component system for
which, as has been mentioned above, crossing point effects can
appear \cite{voll}. Relative volume of each component depends on
magnetic field, current and temperature, but the total volume of
the two components remains unchanged. In this case the crossing
point effect is possible found in this study for $R_{dif}(J,T)$
curves (Fig. 6). Behavior of $R_{dif}(J)$ curves is quite
understandable for a two-component system. With increasing
current, the volume fraction of the normal intergranular medium
increases; whereas, that of the Josephson medium decreases. This
leads to growth in resistance which proceeds up to some current at
which maximal resistance is achieved (Fig. 6). At this current the
volume fraction of the Josephson intergranular medium is zero (or
close to zero). Further increase in current causes a decrease in
resistance since intergranular transport of unpaired
quasi-particle excitations is basically activated in character
(Fig. 1) so that an increase in current enhances the probability
of the intergranular transport. The nature of the second crossing
point at higher current (Fig. 6) is  unclear at the moment and
deserves further study.
\par
In conclusion, we have studied inhomogeneity effects in current
and magnetic-field dependences of the resistive transition to
superconducting state of granular ruthenocuprate
RuSr$_2$(Gd$_{1.5}$Ce$_{0.5})$Cu$_2$O$_{10-\delta}$. In the
current dependence of differential resistance taken at different
temperatures a crossing point effect has been found.
Magnetoresistance in the mixed state of this inhomogeneous system
demonstrates significant hysteresis of different type (clockwise
and anticlockwise). At low temperature the clockwise hysteresis
takes place; whereas, with increasing temperature the transition
to anticlockwise hysteresis occurs. Stepwise structure in MR
hysteretic curves has been observed in low-field range. All the
mentioned phenomena can be explained by taking into consideration
the peculiarities of inhomogeneous state of the granular
superconductor studied, which  consists of superconducting grains
and inhomogeneous intergrain medium.

\section{Acknowledgements}
The authors acknowledge I. Felner (Racah Institute of Physics, The
Hebrew University, Jerusalem) for providing ruthenocuprate
samples. This research was supported by the Robert A. Welch
Foundation (Houston, Texas) under grant A-0514.

\newpage
\centerline{\bf{Figure captions}} \vspace{15pt}

Fig.~1. (Color online) Temperature dependences of the resistance
$R(T)$ of the RuSr$_2$(Gd$_{1.5}$Ce$_{0.5}$)Cu$_2$O$_{10-\delta}$
sample recorded for different measuring currents. The temperature
$T_{c0}=34$~K (shown by arrow) marks a kink in the $R(T)$
dependences and is attributed to the intragrain critical
temperature of transition to the superconducting state. Another
indicated temperature, $T_{cJ}=23.2$~K, is discussed in the main
text. \vspace{15pt}

Fig.~2. (Color online) Resistive superconducting transition of the
sample studied measured at current J = 0.5 mA for different
magnetic fields (in tesla): 0, 0.01; 0.02; 0.03; 0.04; 0.05; 0.06;
0.07; 0.08; 0.09; 0.1; 0.2; 0.3; 0.4; 0.5; 0.6; 0.8; 1.0; 1.2;
1.4; 1.6. Characteristic temperature, $T_{cJ}=23.2$~K, is shown by
arrow.\vspace{15pt}

Fig.~3. Magnetic field dependences of the magnetoresistance,
$[R(H) - R(0)]/R(0) = \Delta R(H)/R(0)$, of the sample studied for
the temperature range $T < T_{c0}$, recorded with increasing field
at current J = 0.5 mA. \vspace{15pt}

Fig.~4. Magnetization curves at $T=10$~K of
RuSr$_2$(Gd$_{1.5}$Ce$_{0.5}$)Cu$_2$O$_{10-\delta}$ sample. The
inset shows $M(T)$ behavior at higher field range. \vspace{15pt}

Fig.~5. (Color online) Temperature evolution of magnetic field
dependences of the magnetization of as-prepared
RuSr$_2$(Gd$_{1.5}$Ce$_{0.5}$)Cu$_2$O$_{10-\delta}$ sample in
low-field range. \vspace{15pt}

Fig.~6. (Color online) Current dependences of the differential
resistance $R_{dif}=dU/dJ$ (obtained from current-voltage
characteristics) at different temperatures. Arrows indicate two
crossing points in the set of $R_{dif}(J)$ curves (at
characteristic currents $J_{x1}$ and $J_{x2}$). \vspace{15pt}

Fig.~7. (Color online) Temperature evolution of MR hysteresis
curves of the sample at $J=3$~mA. It demonstrates the transition
from the clockwise hysteresis at low temperature to the
anticlockwise one with increasing temperature. \vspace{15pt}

Fig.~8. (Color online) Low field behavior of MR hysteresis curves
of the sample at $T=8$~K and $J=3$~mA. \vspace{15pt}

Fig.~9. (Color online) MR hysteresis of the sample at $T=20$~K and
$J=0.5$~mA. Arrow indicates the characteristic field $H_x$ above
which the stepwise structure disappears. The inset presents an
enlarged image of the MR curves in the low-field region.
\vspace{15pt}

Fig.~10. (Color online) The same as in Figure 9 for $T=14.93$~K
and $J=0.2$~mA. \vspace{15pt}

Fig.~11. (Color online) Temperature dependences of characteristic
field $H_x$ and $\Delta R(H) = [R(H)-R(0)]$ ($H=1.6$~T) at
transport currents $J=0.2$ mA and $J=0.5$ mA. Characteristic
temperatures of granular superconductor, $T_{cJ}$ and $T_{c0}$,
are indicated by arrows. The dash-dot horizontal straight line in
the right corner of the upper panel presents $\Delta R(H) =
0$.\vspace{15pt}

Fig.~12. (Color online) Examples of MR hysteresises for $J=0.5$~mA
at temperatures close to the intragrain superconducting
temperature $T_{c0}=34$~K.\vspace{15pt}

Fig.~13. (Color online) Temperature dependences of the relative
amplitude of the resistance jumps for different applied currents.
$R_0$ is zero field resistance. The arrow shows the characteristic
temperature, $T_{cJ}=23.2$~K.

\newpage
\begin{figure}[htb]
\centering\includegraphics[width=0.75\linewidth]{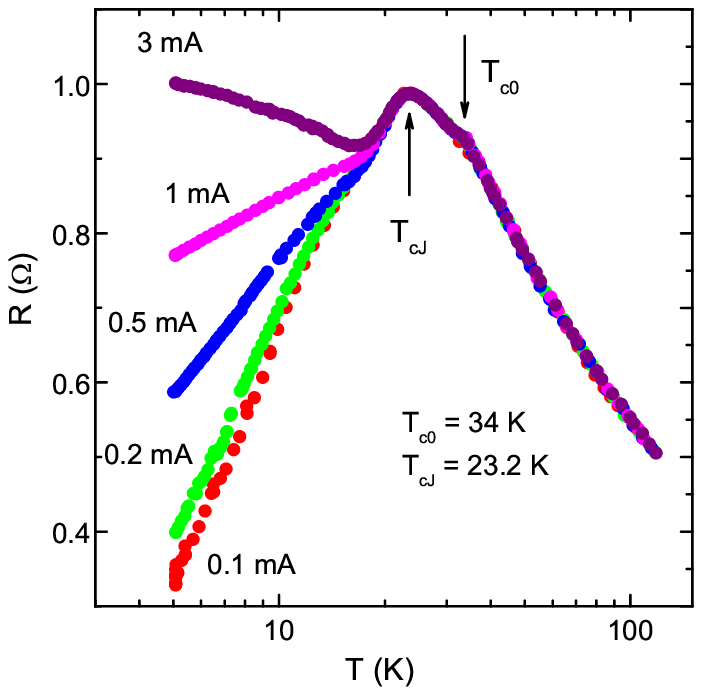}
\centerline{Fig.~1 to paper Belevtsev et al.}
\end{figure}

\begin{figure}[tb]
\centering\includegraphics[width=0.75\linewidth]{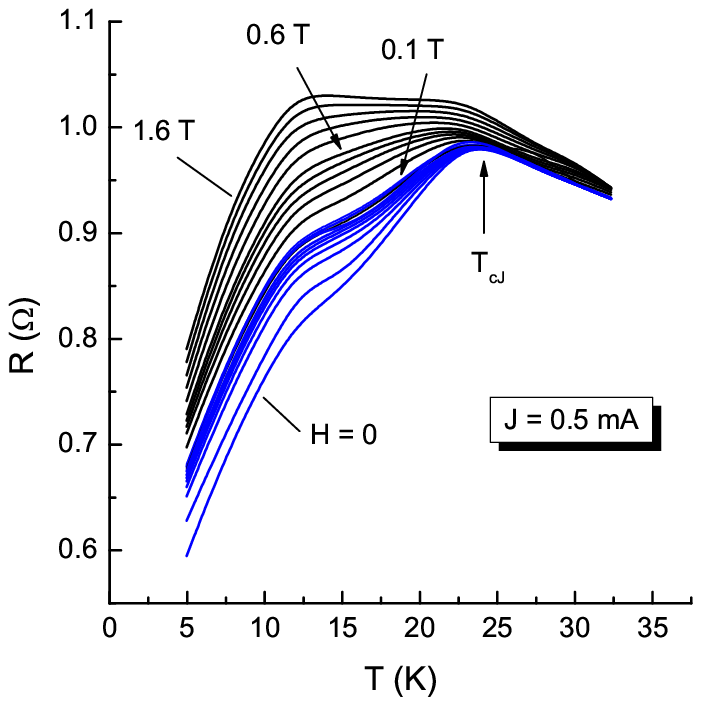}
\centerline{Fig.~2 to paper Belevtsev et al. }
\end{figure}

\begin{figure}[tb]
\centering\includegraphics[width=0.75\linewidth]{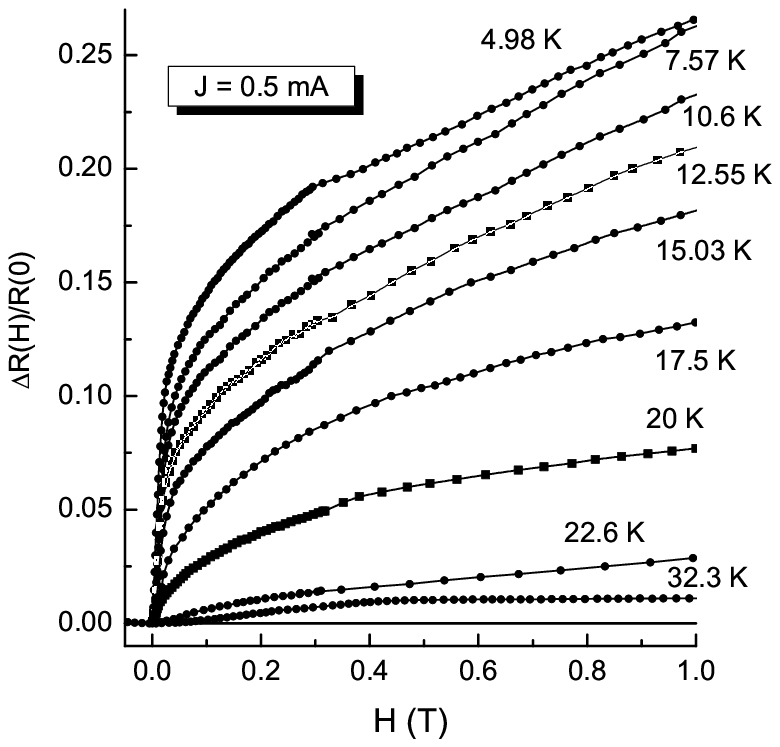}
\centerline{Fig.~3 to paper Belevtsev et al. }
\end{figure}

\begin{figure}[tb]
\centering\includegraphics[width=0.75\linewidth]{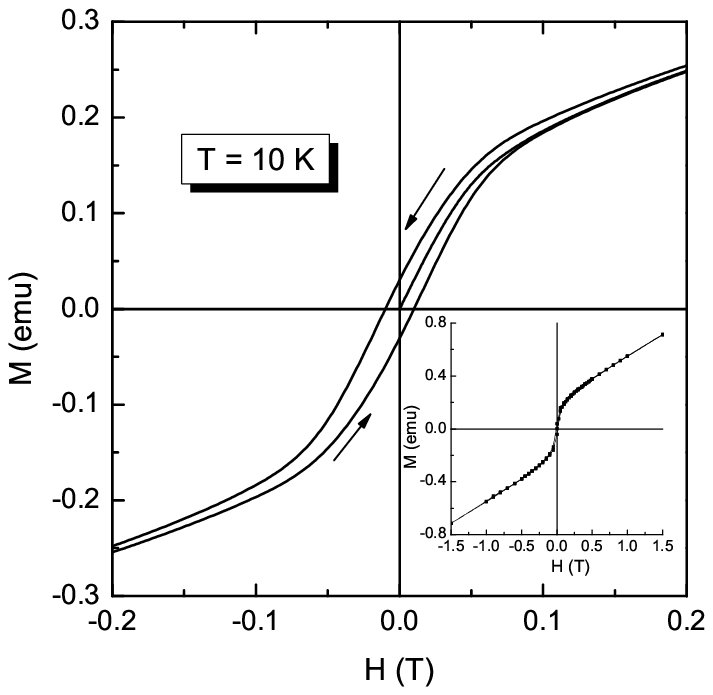}
\centerline{Fig.~4 to paper Belevtsev et al. }
\end{figure}

\begin{figure}[tb]
\centering\includegraphics[width=0.75\linewidth]{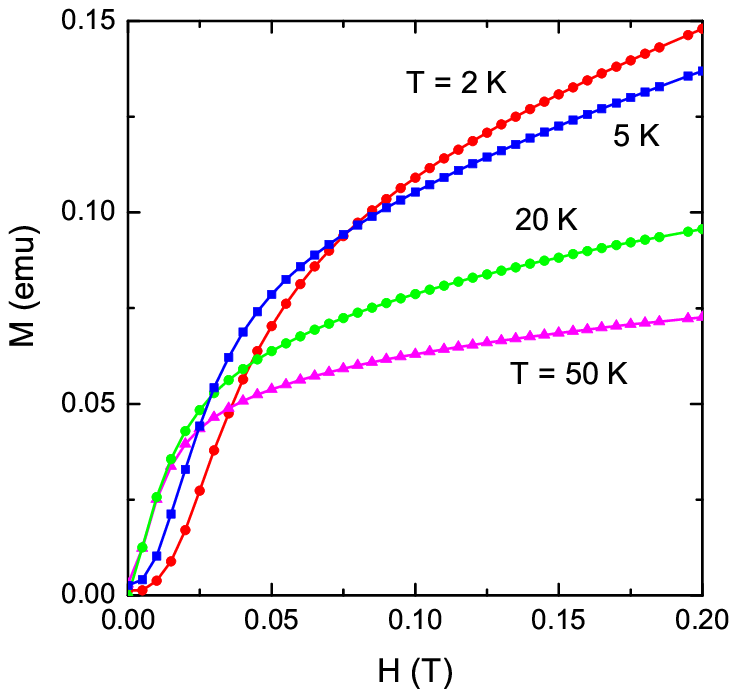}
\centerline{Fig.~5 to paper Belevtsev et al. }
\end{figure}

\begin{figure}[tb]
\centering\includegraphics[width=0.75\linewidth]{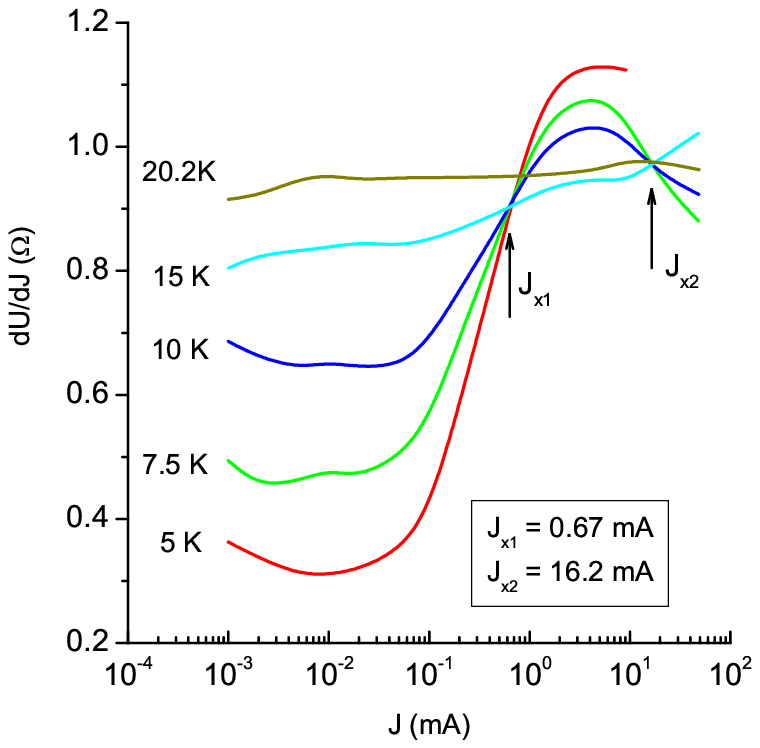}
\centerline{Fig.~6 to paper Belevtsev et al. }
\end{figure}

\begin{figure}[tb]
\centering\includegraphics[width=0.75\linewidth]{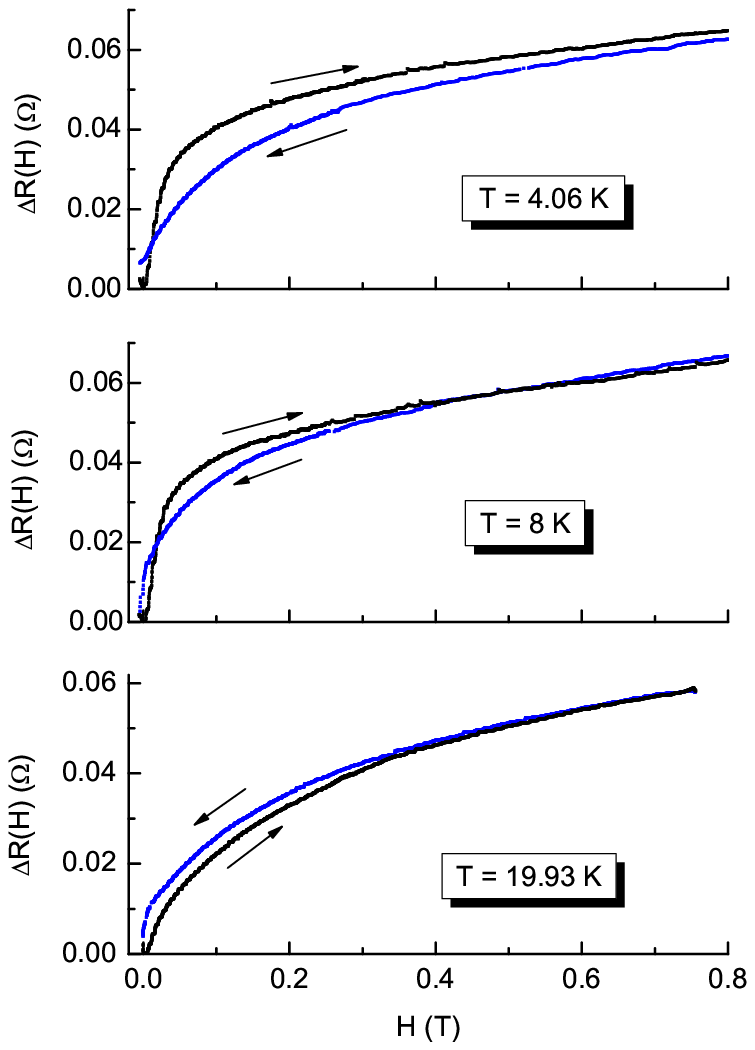}
\centerline{Fig.~7 to paper Belevtsev et al. }
\end{figure}

\begin{figure}[tb]
\centering\includegraphics[width=0.75\linewidth]{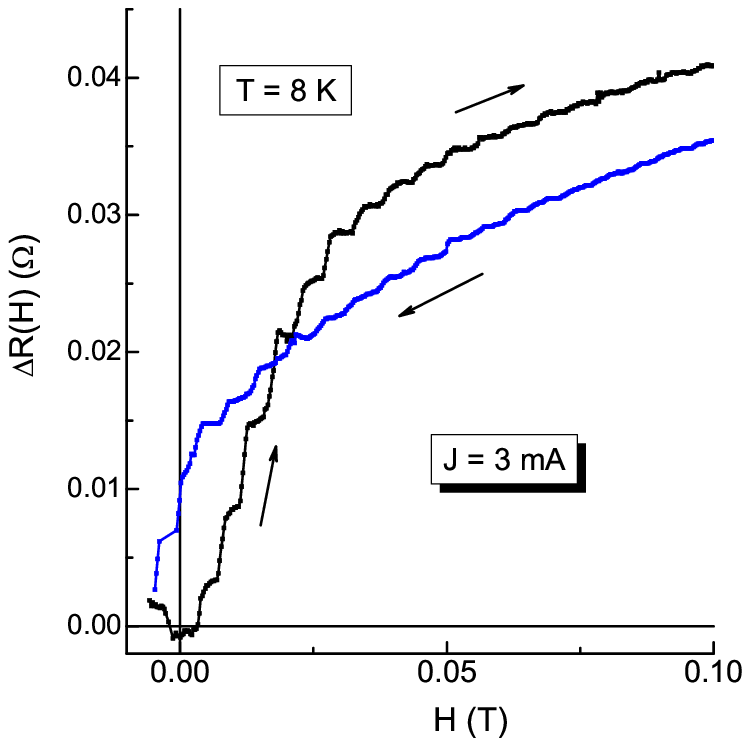}
\centerline{Fig.~8 to paper Belevtsev et al. }
\end{figure}

\begin{figure}[tb]
\centering\includegraphics[width=0.75\linewidth]{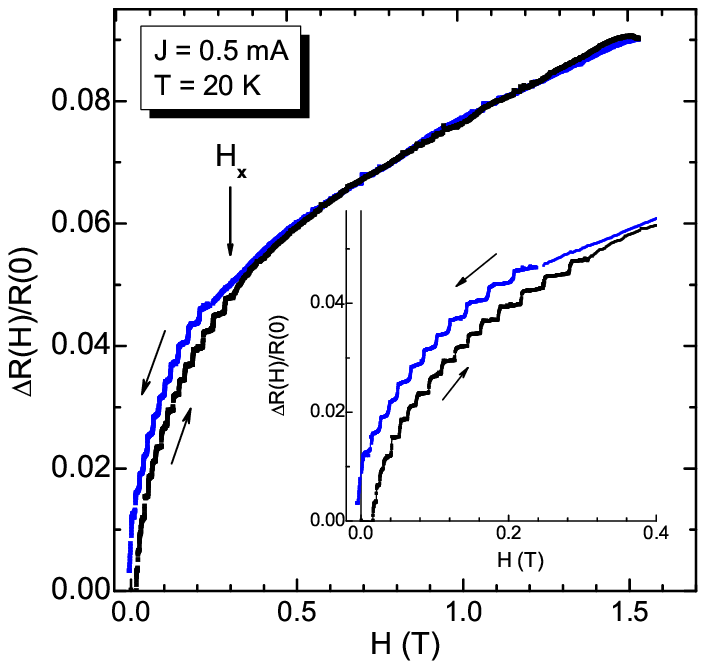}
\centerline{Fig.~9 to paper Belevtsev et al. }
\end{figure}

\begin{figure}[tb]
\centering\includegraphics[width=0.75\linewidth]{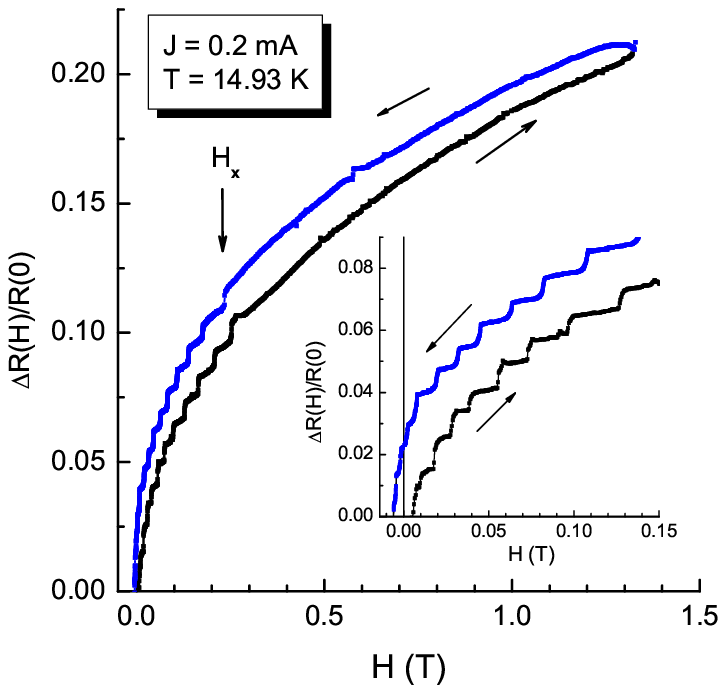}
\centerline{Fig.~10 to paper Belevtsev et al. }
\end{figure}

\begin{figure}[tb]
\centering\includegraphics[width=0.75\linewidth]{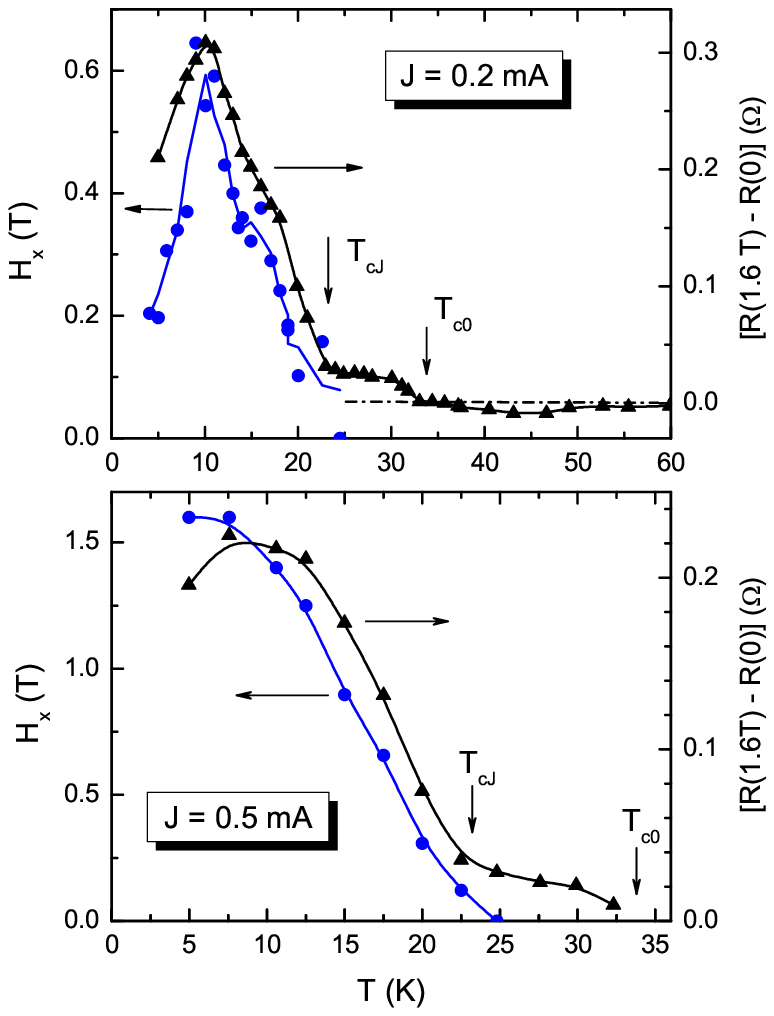}
\centerline{Fig.~11 to paper Belevtsev et al. }
\end{figure}

\begin{figure}[tb]
\centering\includegraphics[width=0.75\linewidth]{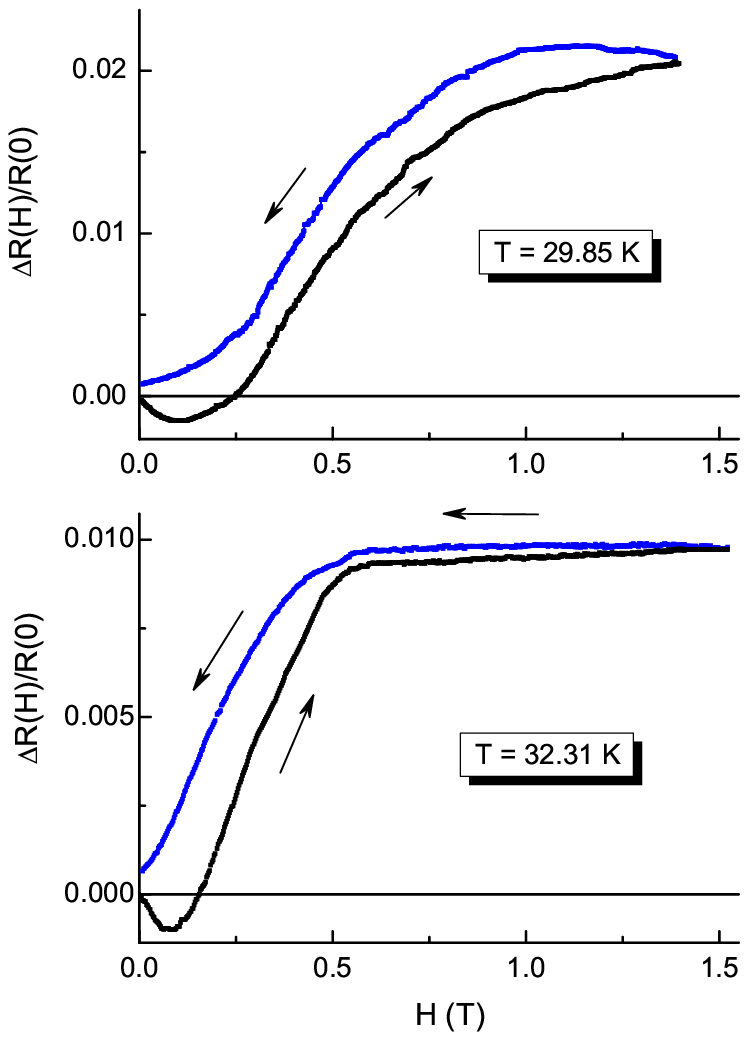}
\centerline{Fig.~12 to paper Belevtsev et al. }
\end{figure}

\begin{figure}[tb]
\centering\includegraphics[width=0.75\linewidth]{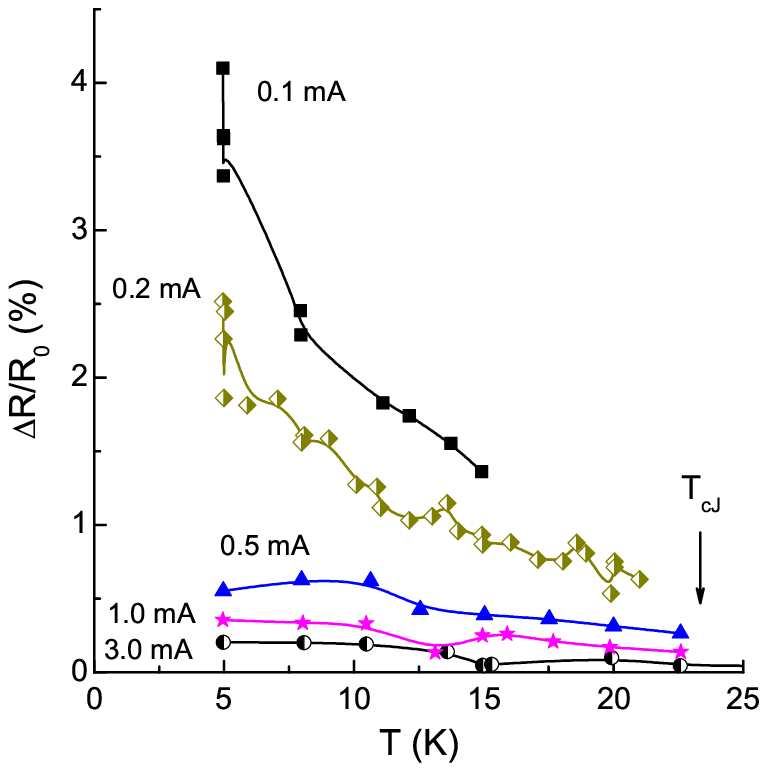}
\centerline{Fig.~13 to paper Belevtsev et al. }
\end{figure}

\end{document}